\documentstyle[twocolumn,prl,aps,psfig]{revtex}
\begin{document}\hbadness=10000
\twocolumn[\hsize\textwidth\columnwidth\hsize\csname %
@twocolumnfalse\endcsname
\title{Free Energy of a Hot Quark-Gluon Plasma\cite{ack}}
\author{Salah Hamieh$^{\rm (a)}$, Jean Letessier$^{\rm (a)}$, 
Johann Rafelski$^{\rm (b)}$, Martin Schroedter$^{\rm (b)}$, 
and Ahmed Tounsi$^{\rm (a)}$}
\address{
$^{\rm (a)}$Laboratoire de Physique Th\'eorique et Hautes Energies,
Universit\'e Paris 7, 2 pl Jussieu, F--75251 Cedex 05\\
$^{\rm (b)}$Department of Physics, University of Arizona, Tucson, AZ 85721\\
}
\date{April 30, 2000}
\maketitle
\begin{abstract}
Thermal (Th-)QCD properties appear to  disagree with lattice (L-)QCD results, 
which suggests that Th-QCD does not have a 
valid perturbative expansion. However, we reproduce L-QCD  using 
the first order $\alpha_s$ corrections in Th-QCD.
\end{abstract}
\pacs{PACS:12.38.Mh, 11.10.Wx, 12.38.Cy, 12.38.Gc}
\vspace{-0.35cm}
]
\begin{narrowtext}
We believe that the sensitivity of thermal (Th-)QCD to the 
value of  QCD coupling,  combined with a common, but
rough approximation of QCD coupling employed
in Ref. \cite{And99}  is the cause of the failure 
to recognize correspondence between lattice (L-)QCD results 
and Th-QCD.  The {\it approximate} two loop result for $\alpha_s^{par}$ used
in the study of quark-gluon liquid\cite{And99} is shown by the dotted 
line in Fig.\,\ref{figrap}. Using  $\alpha_s^{par}$  and comparing to the  pure-glue 
L-QCD results \cite{Boy96}, the authors of Ref.\cite{And99} 
in their Fig.\,1 show  lack 
of agreement between Th-QCD and L-QCD. 
\begin{figure}[tb]
\vspace*{-2.5cm}
\centerline{\hspace*{1.3cm}
\psfig{width=11.6cm,figure=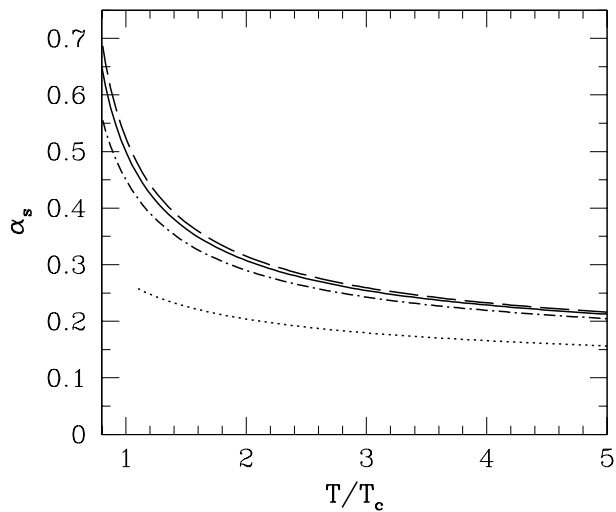}
}
\vspace*{-2.4cm}
\caption{ 
$\alpha_s(2\pi T)$ for $T_c=0.16$\,GeV. 
Dashed line: $\alpha_s(M_Z)$=0.119; solid line =0.118;
dot-dashed line =0.1156.
Dotted line: approximation used by 
Ref.\protect\cite{And99}.
 \protect\label{figrap} 
}
\vspace*{-.2cm}
\end{figure}

As the large coefficients in free energy ${\cal F}$ 
weak coupling expansion  suggest\cite{And99,Bra96,Kas95,Arn94}, 
the convergence of Th-QCD is not expected in any scheme considered
so far. For this reason it seems to us to be more appropriate to 
check only the lowest order of this possibly semi-convergent series. 
In Fig.\,\ref{figP}, we show  a comparison of the first order 
in $\alpha_s$ Th-QCD result for pressure $P/T^4=-{\cal F}/VT^4$,  with  
a L-QCD calculation which includes dynamical quarks\cite{Kar00}.  
To obtain these results:
a) we use numerically computed $\alpha_s(\mu=2\pi T\equiv\kappa T/T_c)$, 
solid line in  Fig.\,\ref{figrap}, obtained integrating 
the renormalization
group equation \cite{LTR96}, using the two loop $\beta$-function;
b) we assume that $T_c\simeq$150--160\,MeV;
and thus $\alpha_s(2\pi T)\simeq\alpha_s(1\,{\rm GeV}\cdot T/T_c)$\,;
c) we allow for a vacuum property and add to the quark-gluon
free energy the vacuum term $ 
{\cal F}_V=BV,\,B=0.19\,{\mbox{GeV}}/{\mbox{fm}}^3$, this influences
{\it only} the slope in $P$ near to $T=T_c$;
d) we took for the case 2+1 flavors a renormalized 
strange quark mass $m_s/T=1.7m_s^0/T=1.7$, which leads to
a 50\% reduction in strange quark number.

\begin{figure}[tb]
\vspace*{-2.95cm}
\centerline{\hspace*{.8cm}
\psfig{width=11.7cm,figure=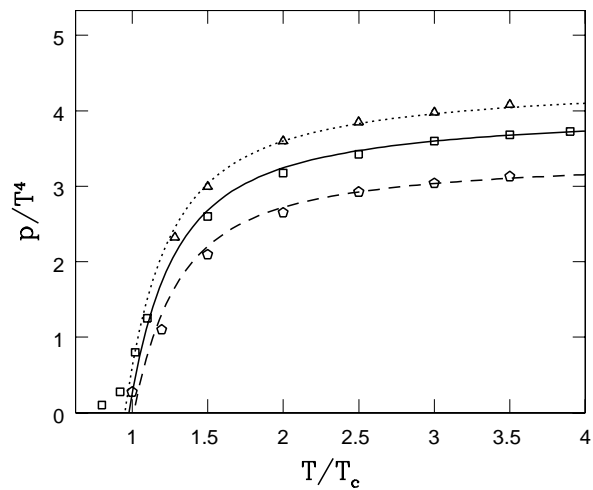}
}
\vspace*{-2.4cm}
\caption{ 
L-QCD results \protect\cite{Kar00} for $P/T^4$ compared 
with our semi-perturbative approach: 
dotted line 3 flavors, solid line 2+1 flavors, 
and dashed line 2 flavors.
 \protect\label{figP} 
}
\vspace*{-.2cm}
\end{figure}

We conclude that for $T\ge 2.5T_c$, ${\cal O}(\alpha_s)$-corrections
accurately describe L-QCD free energy of a 
hot quark-gluon plasma. Introducing 
a  vacuum energy-pressure $B$, L-QCD results are 
reproduced  near to $T=T_c$.


\vskip -0.7cm

\end{narrowtext}\end{document}